# Feature Decomposition Based Saliency Detection in Electron Cryo-Tomograms

Bo Zhou[1], Qiang Guo[2], Xiangrui Zeng[3], and Min Xu[3*]

[1]Robotics Institute, Carnegie Mellon University, Pittsburgh, USA
[2]Max Planck Institute for Biochemistry, Martinsried, Germany
[3]Computational Biology Department, Carnegie Mellon University, Pittsburgh, USA
[*]Corresponding author

## Abstract

Electron Cryo-Tomography (ECT) allows 3D visualization of subcellular structures at the submolecular resolution in close to the native state. However, due to the high degree of structural complexity and imaging limits, the automatic segmentation of cellular components from ECT images is very difficult. To complement and speed up existing segmentation methods, it is desirable to develop a generic cell component segmentation method that is 1) not specific to particular types of cellular components, 2) able to segment unknown cellular components, 3) fully unsupervised and does not rely on the availability of training data. As an important step towards this goal, in this paper, we propose a saliency detection method that computes the likelihood that a subregion in a tomogram stands out from the background. Our method consists of four steps: supervoxel over-segmentation, feature extraction, feature matrix decomposition, and computation of saliency. The method produces a distribution map that represents the regions' saliency in tomograms. Our experiments show that our method can successfully label most salient regions detected by a human observer, and able to filter out regions not containing cellular components. Therefore, our method can remove the majority of the background region, and significantly speed up the subsequent processing of segmentation and recognition of cellular components captured by ECT.

Keywords: saliency detection, Electron Cryo-Tomography, super-voxel segmentation, 3D Gabor filter, robust PCA

## 1 Introduction

The development of Electron Cryo-Tomography (ECT) has enabled the detailed inspection and visualization of subcellular structures at sub-molecular resolution and in near-native state[19, 24]. With this cellular imaging technique, subcellular components can be systematically analyzed at unprecedented level of detail and faithfulness. Recent studies have shown this *in situ* visualization enables the discovery of numerous important structural features in complex virus[10, 11], as well as in prokaryotic and eukaryotic cells[4, 9, 14]. ECT has established its position as one of the leading techniques for visualizing the subcellular and macromolecular organization of single cells[5].

In principle, an ECT tomogram captures structural information of all cellular components in the field of view. However, several factors, such as the low signal-to-noise ratio (SNR), the limited tilt projection range (missing wedge effect) and the crowded nature of intracellular structures, make the systematic structural analysis of cellular components captured by ECT very difficult. In addition, tomograms are large size gray scale 3D images. A typical raw tomogram can have a size of $6000 \times 6000 \times 1500$ voxels, which is very computationally expensive to process. In order to analyze the cellular components captured by ECT tomograms, these cellular components must first be segmented and recognized from these tomograms. Currently, most of the segmentation is performed either manually, or automatically. The manual segmentation of cellular components in such 3D images is very laborious, even with the aid of computational segmentation tools like watershed transform and thresholding [25]. On the other hand, most of the automatic segmentation is very computationally intensive and often limited to specific types of objects. Many of these automatic segmentation methods rely on matching of the geometrical model of the specific cellular structure, such as filaments, microtubules and membranes [18, 23, 16]. Recently, supervised methods have been developed that segments specific cellular structures using classification models trained on annotated training images of the specific cellular structures of interest [8, 17]. Such supervised segmentation methods are restricted to segmenting specific structures in tomograms that are already characterized by human annotator.

A generic segmentation algorithm that is able to segment general cellular components including unknown cellular components, without considering the availability of training data is needed to complement the existing segmentation methods. As an important step towards this goal, we propose a saliency detection method and address the problem of automatically extracting the salient region from an unknown background in ECT. The *saliency* of an image subregion is the likelihood that it stands out relative to its background. Given that the perceptual, cognitive and computational resources are limited, to facilitate such segmentation, it is important to ignore the background regions which occupy the majority portion of the tomogram.

To solve the saliency detection problem, we propose a method that consists of following steps: supervoxel over-segmentation, feature extraction, feature matrix decomposition, and computation of saliency. The method produces a map that represents the saliency of subregions of a tomogram. Our experiments show that 1) our method can successfully perform saliency detection compared with human annotation, and 2) the number of detected salient voxels is significantly smaller than the total number of voxels. As a result, our approach can substantially reduce the background region, and accelerate the succeeding segmentation and analysis of the cellular components in the ECT.

Our main contributions are summarized as follows:

- We design an easy-to-implement salient region detection method based on feature decomposition.

- With the candidate regions of cellular objects in ECT, the computational cost of tasks like object detection and segmentation of specific structure is greatly reduced, facilitating the isolation of different cellular structure of interest.

## 2 Methods

Our method comprises five major steps, including data pre-processing for de-noising, supervoxel over-segmentation, feature extraction, feature matrix construction, and computation of saliency. The flow diagram of the method is shown in Figure 1. The 1st stage is intended to enhance the image contrast to improve the performance of supervoxel over-segmentation in the 2nd stage, as well as the quality of the extracted feature in the 3rd stage. The different stages are described in detail in the following sections. The procedure takes the intuitive assumption that, in terms of image features, the salient regions are the minority regions in the tomograms.

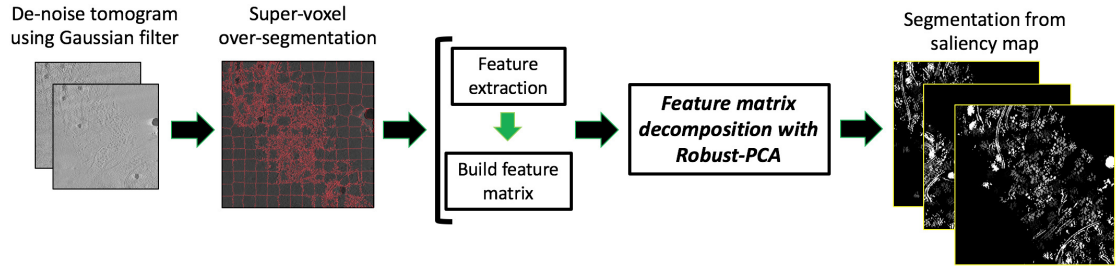

Figure 1: General flow diagram of the method for salient object detection and segmentation.

The details of the de-noising stage are described in Appendix 6.1. In this paper, we assume that all the salient objects have a size bigger than the assigned kernel scale. The scale $\sigma$ is chosen by visually evaluating the processed data using different scales. Given the majority of noise is smaller than the scale selected, this step can filter out the noise and preserve the useful information in the tomogram. We chose $\sigma = 1.8$ for all experiment in this paper.

## 2.1 Supervoxel over-segmentation

The geometrical features are important information for analyzing the tomograms. Geometry is the primary feature to be considered when researchers choose the interested region in a tomogram to analyze. In this step, we use a supervoxel over-segmentation method, simple linear iterative clustering (SLIC)[3], to over-segment the tomogram into sub-volumes which generate 3D geometric boundaries and clusters that enclose small volumes with similar density within a certain distance of the neighborhood. In ECT tomogram, low gray-scale levels represent the electron-dense region.

We generalize the method in R. Achanta's paper[3] to three dimensions (x, y, z) and gray-scale (I). Each voxel corresponds to a vector in $\mathbb{R}^4$. The vectors are used for clustering. One of the method's parameter is $n$, the desired number of equally sized supervoxels.

The clustering procedure begins by initializing $n$ cluster centers at points spaced $S$ voxels apart on a regular grid, where: $S = \sqrt{\frac{N}{n}}$, and $N$ is the number of the voxels in the tomogram. There are approximately $S$ voxels in each supervoxel. Then each voxel is assigned to the nearest cluster center whose search region overlaps its spatial location. The raw distance measurement $D'$ calculates the distance between a voxel and cluster center. $D'$ is defined as

$$D' = \sqrt{w_I(\frac{d_I}{m})^2 + w_s(\frac{d_s}{S})^2} \qquad (1)$$

, where spatial distance

$$d_s = \sqrt{(x_c - x_i)^2 + (y_c - y_i)^2 + (z_c - z_i)^2} \qquad (2)$$

and intensity distance

$$d_I = \sqrt{(I_c - I_i)^2} \qquad (3)$$

. $D'$ is further simplified as

$$D = \sqrt{d_I{}^2 + w(\frac{d_s}{S})^2 m^2} \qquad (4)$$

, where weight coefficient $w$ is the second parameter in the 3D SLIC clustering, the compactness of clusters. The voxel clusters for tomograms can be generated by iterating every voxel in it. In short, there are two parameters we need to specify, the number of clusters $n$ and the compactness of cluster $w$. A higher value of $n$ and lower value of $w$ should be chosen if the potential content is small and morphological complex, which can produce higher spatial density cluster and more deformable shape to enclose the region containing potential objects, vice versa.

## 2.2 Feature extraction

After the supervoxel over-segmentation, for each supervoxel, we perform feature extraction and construct a feature vector with a size of 30. Two groups of the features are calculated:

### 2.2.1 3D Gabor filter based features

Previous studies have shown that the Gabor function provides a useful and reasonably accurate description of most spatial aspects of simple receptive fields[15].

Gabor function is a product of a Gaussian kernel and a complex sinusoid function. The derivation of extending 2D to 3D Gabor function is shown in Appendix 6.2. We assume the Gaussian envelop have the same scale in three axes; Thus the shape of this Gaussian can only be spherical but can have different scales. By modifying sinusoid frequency in three orientations and scale parameters, we generate 24 3D Gabor filters by rotating 0, 45, 90 and 135 degrees about the x, y, and z-axes, as well as two different Gaussian scales. Each Gabor feature is calculated by taking the mean of the filter response inside the supervoxels.

### 2.2.2 Density features

A tomogram is reconstructed from multiple electron beam projections at different angles. Higher density material causes more energy attenuation of the electron beam when arriving at the energy detector. This results in a lower voxel intensity corresponding to a denser cellular structure. Cellular structure density features are inversely related to the intensity value in the tomogram. Given this information, the second group of features consists of the intensity distribution in the supervoxel. The intensity's dynamic range could vary between ECT tomograms due to different settings of the imaging system and reconstruction parameters. With that, we first normalize the dynamic range into [0 600] and generate a six bins histogram range from the 0 to 600. The number of voxels counted in each bin is used as one density feature. There are six density features extracted from the histogram of each supervoxel.

After segmenting the tomogram into candidate supervoxels, we compute the feature vector for each supervoxel, and construct a feature matrix with each row corresponding to the feature vector of a supervoxel. The feature matrix with the size of $n$ by 30 will be used in the next step, where $n$ is the number of supervoxels.

## 2.3 Feature matrix decomposition

Previous studies have shown that subspace estimation by sparse representation and rank minimization is an excellent unsupervised method for separating the salient image regions from the image background [6, 22]. We decompose the feature matrix constructed from the last section into a low-rank matrix and a sparse matrix using robust principal component analysis (RPCA) via principal component pursuit (PCP)[7]. Each row of the sparse matrix represents the saliency of individual supervoxels, whereas the low-rank matrix represents the common background information. The feature matrix $F$ is represented as $F = L + S$, where $L$ is a low-rank matrix, and $||S||_0$ is the L0 norm of the matrix $S$. The minimization of $||S||_0$ enforces $S$ to be a sparse matrix with a small fraction of nonzero entries. The decomposition is an optimization process in the following form:

$$\min_{L,S} \ rank(L) + \beta||S||_0 \qquad s.t. \quad F - L - S = 0 \qquad (5)$$

, where $\beta \geq 0$ is a hyperparameter. The minimization of Equation 5 is NP-hard. Instead of directly solving Equation 5, PCP transforms Equation 5 to an equivalent convex optimization problem[7]:

$$\min_{L,S} \ ||L||_* + \beta||S||_1 \qquad s.t. \quad F - L - S = 0 \qquad (6)$$

, where $||L||_*$ is the nuclear norm of $L$, calculated from the singular values of $L$, $||S||_1$ is the L1 norm of $S$. The RPCA-PCP method can effectively recover the low-rank and the sparse matrix for our feature matrix by optimizing Equation 6 [7]. In an ideal decomposition of the feature space, most of the image background should lie in a low dimensional subspace so that they can be represented as a low-rank matrix $L$ shown above.

## 2.4 Calculation of saliency map and salient region segmentation

Given the sparse matrix $S$ decomposed from the feature matrix $F$, each supervoxel's saliency can be represented by corresponding row in the sparse matrix $S$. In this section, we refer it as saliency vector of supervoxel. After obtaining the saliency vector of the corresponding supervoxel, we calculate the representation of saliency by taking the mean of the saliency vector, and the saliency is assigned to the corresponding supervoxel's spatial location which generates a saliency volume.

## 2.5 Evaluation of salient region segmentation

### 2.5.1 Obtaining ground truth through manual annotation

Ground truth salient region annotation is obtained by using the human annotator selected supervoxel regions where the annotator believe the regions stand out from background. The anchor points are dropped by the annotator to select regions of such supervoxels. The annotator has no previous knowledge about the salient map and any output produced by our method.

### 2.5.2 ROC based performance measure

Binary salient region segmentation of the saliency region is generated by setting a threshold value to the saliency map volume (Figure 3). The performance of the segmentation using the saliency volume is evaluated using receptor operating curve (ROC) analysis[21]. A true positive in our ROC analysis is if the manually annotated anchor point falls within the salient region segmentation. Three tomogram slices in three different datasets are evaluated. The mean area under the curve (AUC) are also calculated for each dataset (Figure 4).

## 3 Results

Three different experimental tomograms are included to test the method's performance. Tomogram 1 and tomogram 3 are obtained from the EMPIAR public image archive [13]. Tomogram 1 (EMPIAR ID: 10045) captures a sub-region of the purified S. cerevisiae ribosomes[2]. Tomogram 2 contains a sub-region of a rat's primary neuron, which was collected at Max Planck Institute of Biochemistry[12]. Tomogram 3 (EMPIAR ID: 10048) captures a sub-region of the Chlamydia trachomatis secretion system[1].

### 3.1 Supervoxel over-segmentation via SLIC clustering

The optimal scale space for different tomograms can vary due to the noise introduced by different imaging data acquisition parameters and reconstruction methods. Using a fixed imaging and reconstruction protocol, an optimal scale can be chosen. A de-noised ECT tomogram using different scale 3D Gaussian filters are shown in Appendix 6.1 and Figure A.1. In this tomogram, we choose scale $\sigma = 1.8$ voxels, which best preserves the cellular structural information and rule out noise. Bigger or smaller filter scale will over-blur or under-de-noise the tomogram. The tomogram de-noised with optimal Gaussian filter is used for all results shown later.

Using the iterative clustering method discussed in Section 2.1, we cluster a tomogram into different number of supervoxel. The compactness of the supervoxel is set by $w$, which means bigger $w$ will constrain the deformation of supervoxel. Figure 2 shows the supervoxel over-segmentation using the different combination of $n$ and $w$.

In principle, to avoid missing small and complex salient region in tomogram, we choose a bigger number of supervoxel $n$ and smaller compactness coefficient $w$ to make sure these complex regions are clustered and enclosed in supervoxels so that the saliency can be detected in later steps. We choose $n = 10000$ and $w = 0.025$ as the optimal parameters for SLIC clustering in the three test tomograms.

### 3.2 Extraction of feature and construction of saliency map

By applying the 3D Gabor filters with different orientations to the tomogram, we extract the corresponding feature in each voxel. Given the supervoxels' spatial information, voxel's 3D Gabor features are assigned to the corresponding supervoxel. Example of feature extraction results with different 3D Gabor filters is shown in Appendix 6.2 and Figure B.1. The supervoxel's density features are extracted by generating the six bins histogram for each supervoxel; each bin represents one density feature. The cellular structure, like ribosome, membrane, microtubule usually have higher density, so the first three density features of the region contains such cellular structure will have higher values. Whereas, the last three density features will be high if supervoxel contains low-density cellular structures.

Given the feature vectors of every supervoxel in the tomogram, we construct the feature matrix and apply RPCA to decompose it into the low-rank matrix and the saliency matrix. The saliency of supervoxel is calculated using the rows in the saliency matrix and assigned to the spatial location of the supervoxels.

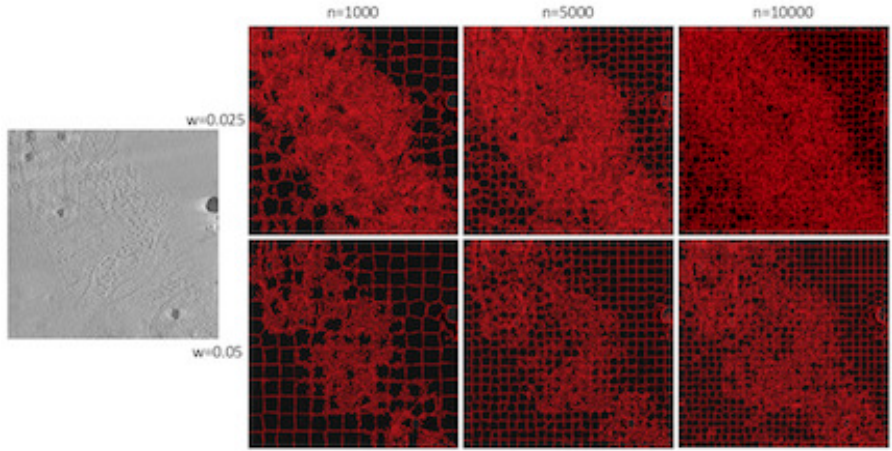

Figure 2: SLIC Supervoxel over-segmentation in tomogram 2 with different number of supervoxel and compactness setting. First column: one slice from the de-noised tomogram. Second to Fourth column: visualization of supervoxels' margins in this slice with $n$=1000, 5000, 10000 (columns) and $w$=0.025, 0.05 (rows).

### 3.3 Evaluation of saliency detection

Our method is tested and evaluated in three different ECT tomograms consisted of simple and complex cellular contents. Figure 3 shows the saliency results generated using the tomograms from these three tomograms. The structural composition of tomogram 1 and tomogram 3 are relatively simple. As we can see, our method can efficiently detect the saliency regions. Tomogram 2 contains cellular structures with various size, density, and shape. More details of salient region visualization are shown in Appendix 6.3. Our method can effectively represent the saliency of different subregions enriched with compact cellular structures in this tomogram.

The performance of salient region segmentation using saliency map is evaluated using the three tomograms mentioned above. Three slices with ground truth anchor annotation in these three tomograms are used to generate the ROC analysis. ROC curve for these tomogram slices is shown in Figure 4. The average AUC for tomogram 1 and tomogram 2 are approximately equal to 0.89 and 0.863. The average AUC for tomogram 3 is about 0.815. The performance of salient region segmentation is better in the tomogram 1, compared to segmentation in tomogram 2 and 3, due to its simple and uniform structural content.

The number of detected salient supervoxels and voxels are recorded using the optimal cutoff threshold obtain from the ROC experiment shown above. The results are shown in Table 1. The number of salient supervoxel and salient voxels are significantly smaller than the total number of supervoxels and voxels. The background voxels are filtered out using this method. Therefore the automatic segmentation and consequent processing steps can be significantly sped up by only processing such a small number of salient voxels.

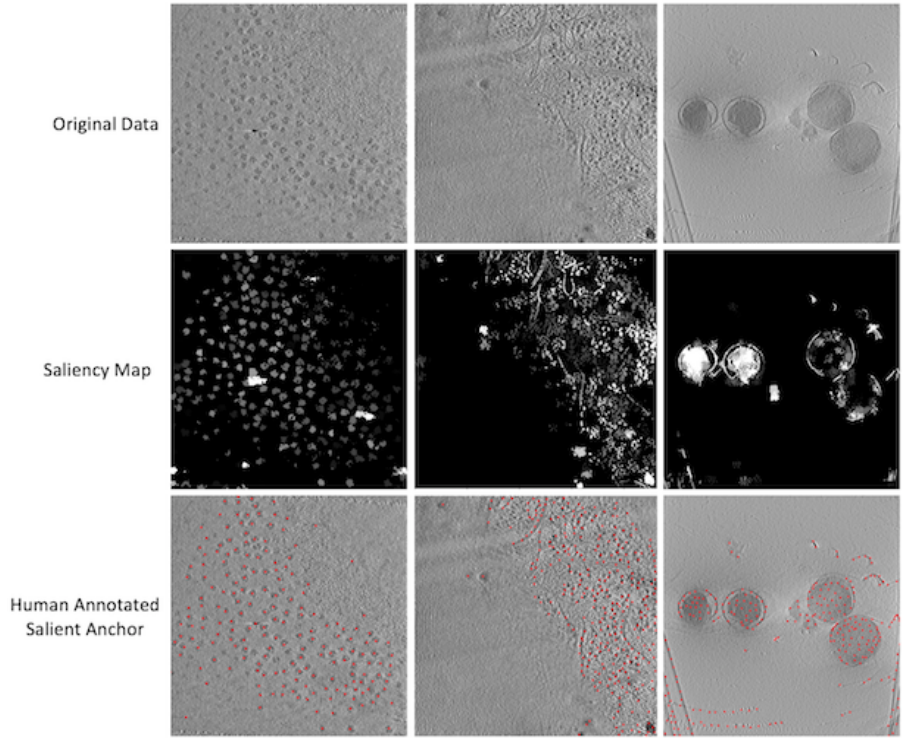

Figure 3: Visualization of example slices of saliency map (saliency level below $\frac{1}{2}(S_{max}+S_{min})$ set to 0) along with corresponding tomogram slices and human annotated salient region anchor in 3 different tomograms.

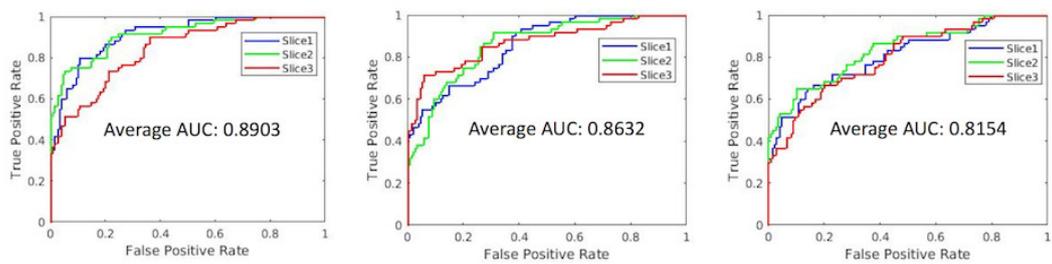

Figure 4: Segmentation evaluation with ROC curve using saliency volume.

Table 1: Salient supervoxel selected from tomogram

|  | Tomogram 1 | Tomogram 2 | Tomogram 3 |
| --- | --- | --- | --- |
| Number of Supervoxel | 10000 | 10000 | 10000 |
| Number of Salient Supervoxel | 869 | 1324 | 330 |
| Percentage of Selected Supervoxel | 8.69% | 13.24% | 3.30% |
| Number of voxel | 321,553,500 | 134,522,880 | 37,796,240 |
| Number of Salient voxel | 25,768,154 | 14,756,210 | 1,564,276 |
| Percentage of Selected voxel | 8.01% | 10.96% | 4.14% |

# 4 Discussion

Efficient and fully unsupervised automatic segmentation of all cellular components in a tomogram is extremely challenging, because of the high degree of structural complexity and the imaging limitations in the cellular electron cryo-tomograms. As an important step towards this goal, we have proposed an efficient salient region detection method in ECT that mimics the human visual system for detecting a tomogram's salient sub-regions that contain cellular components. The quantitative results obtained suggest that our method is useful in salient region detection in studies using ECT technique.

A direct application of our saliency detection is for template free particle picking in ECT tomograms. A tomogram usually contain macromolecules of diverse structures of diverse size. Currently, the main computational approach for selecting such diverse macromolecules without the use of a structural template is through Difference-of-Gaussian(DoG) filtering based template-free particle picking [26, 20]. However, such approach tend to select macromolecules with globular structure and with certain size range. By contrast, the connected salient supervoxels in the binarized saliency map generated from our approach can directly serve as candidate macromolecules or other subcellular components without structure and size constrains. Therefore our saliency detection approach can potentially be used as a more powerful and flexible template-free particle selector.

Future efforts to combine our method with the automatic analysis system for systematically analyzing the object using adapted structural pattern mining methods [e.g., 29, 28, 30]. Such combination will allow us to tremendously reduce the time for laborious manual annotation of tomogram and discover the underlying correlation between structures.

# 5 Acknowledgements

We thank Dr. Robert F. Murphy for suggestions. This work was supported in part by U.S. National Institutes of Health (NIH) grant P41 GM103712. MX acknowledge support from Samuel and Emma Winters Foundation.

# 6 Appendix

## 6.1 De-noising

The noise in a tomogram can directly impact every subsequent step in our method. Thus, de-noising and preserving useful information for the input tomogram is critical. According to the scale-space theory, we can isolate information according to the spatial scale. Given a scale, all the features with a size smaller than the designed scale can be filtered out, whereas the others are preserved[27]. In this step, we implement the scale-space theory with a sampled 3D Gaussian kernel, which aims for isolating useful information from noise by directly convolving the Gaussian kernel with tomogram. In three dimension, continuous 3D Gaussian kernel with scale $\sigma$ is defined by $G[x;\sigma] = \frac{1}{\sigma^3(2\pi)^{\frac{3}{2}}} exp(-\frac{x_1^2+x_2^2+x_3^2}{2\sigma^2})$. The tomogram $T$ is then convolved with G, and generates the de-noised volume $V$: $V[x;\sigma] = G[x;\sigma] * T$

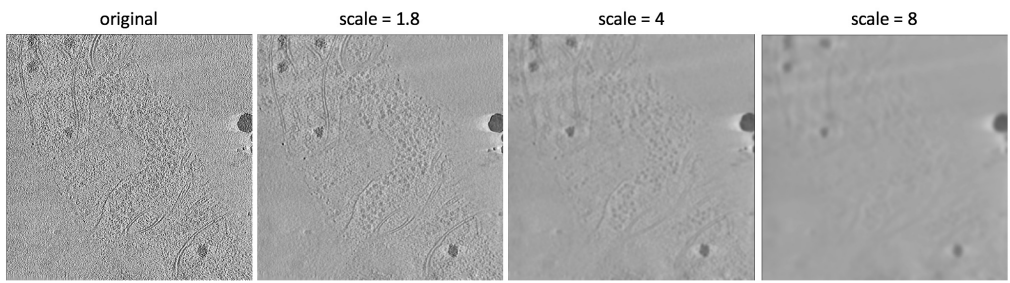

Figure A.1: Gaussian filtered tomogram with different scale 3D Gaussian filter. Three de-noised tomograms are produced by convolving different scale 3D Gaussian filter to the original tomogram. De-noised tomogram with scale $\sigma=1.8$ better preserve detailed structural information compared to de-noised tomogram with scale $\sigma=4$ and scale $\sigma=8$.

## 6.2 3D Gabor filter and feature

Here, we extend the 2-D Gabor function to a 3-D function. The 3D Gabor function (H) can be written as $H[x,y,z] = G[x,y,z] * S[x,y,z]$, where $G$ is a 3D Gaussian envelope $G[x,y,z;\sigma] = \frac{1}{\sigma^3(2\pi)^{\frac{3}{2}}} \exp(-\frac{x^2+y^2+z^2}{2\sigma^2})$, and $S$ is a complex sinusoid function $S[x,y,z] = \exp[2\pi i(Ux+Vy+Wz)]$, where $U$, $V$, and $W$ are the 3D frequencies of the complex sinusoid. They determine Gabor filter's orientation and spacing in the spatial domain.

## 6.3 Visualization of salient region with saliency

Examples of the salient region with the corresponding saliency in different slices of the tomogram that captures a sub-region of rat's primary neuron are shown in Figure C.1. It can be seen that the regions containing bigger cellular structure with higher material density normally present higher saliency, compared to the structure with smaller size, simpler structure, and lower density. This behavior of our method matches the human behavior when observes imaging data. Human observer normally first detect big object with higher density and complex shape, and then simpler and smaller object.

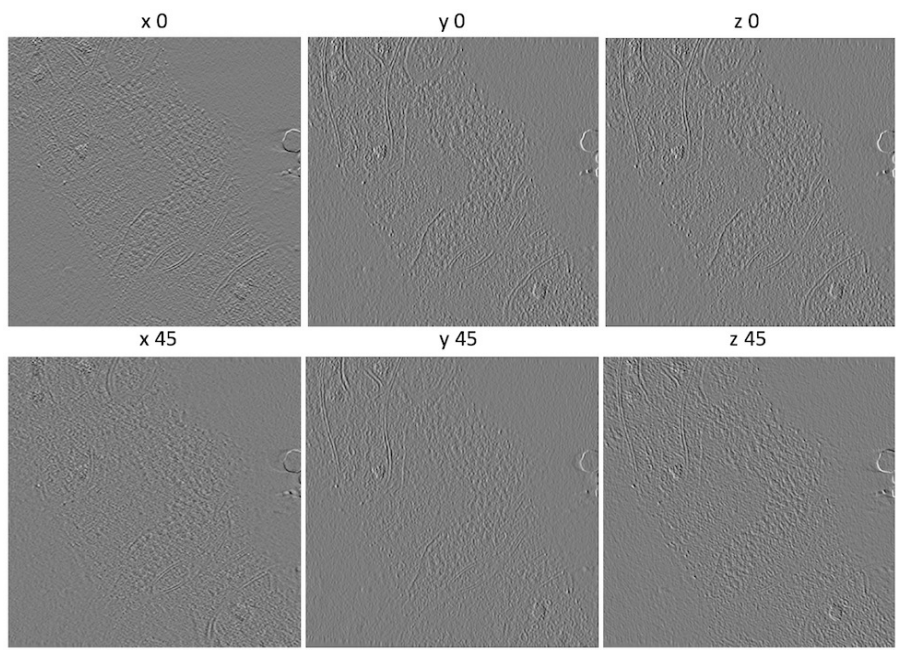

Figure B.1: Visualization of extracted features from tomogram by applying different 3D Gabor filters. First row: Gabor response in x, y and z direction with 0 degree of rotation along the axes. Second row: Gabor response in x, y and z direction with 45 degree of rotation along the axes.

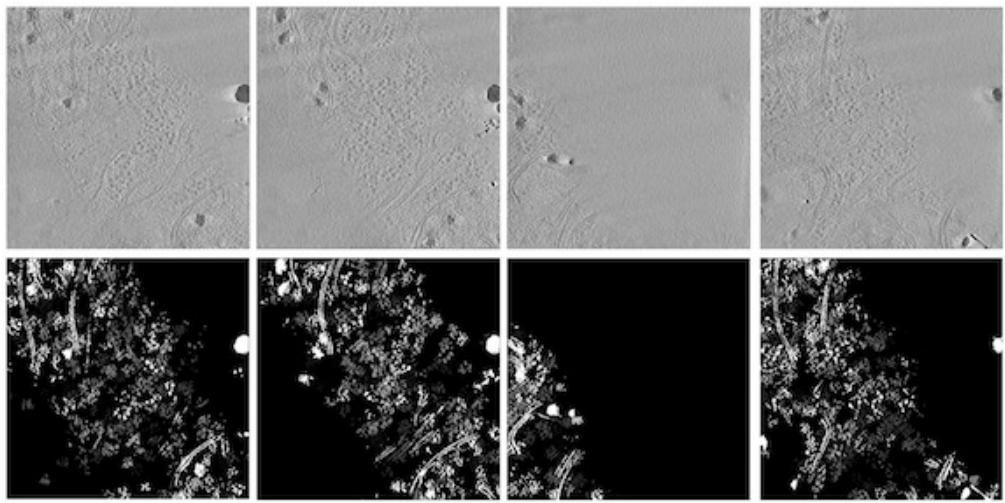

Figure C.1: Visualization of saliency map (second row) along with the corresponding tomogram slices (first row). saliency level below the $\frac{1}{2}(S_{max} + S_{min})$ is set to 0 in this visualization. Different regions present different level of saliency due to the various complexity of content.